\documentclass[aps, amsmath,amssymb, reprint, prx]{revtex4-2}

\usepackage{graphicx}% Include figure files
\usepackage{dcolumn}% Align table columns on decimal point
\usepackage{bm}% bold math
\usepackage{xcolor}
\usepackage[mathlines]{lineno}% Enable numbering of text and display math
\begin{document}

\title{Fundamental limitations of thermoradiative energy conversion}

\author{Maxime Giteau}
 \email{maxime.giteau@cnrs.fr}
 \affiliation{CNRS, Laboratoire PROcédés, Matériaux et Energie Solaire (PROMES), UPR 8521, Odeillo, France}
 \affiliation{ICFO - The Institute of Photonics Sciences, Castelldefels, Barcelona, Spain}
\author{Michela F. Picardi}%
\author{Georgia T. Papadakis}%
  \email{georgia.papadakis@icfo.eu}
\affiliation{ICFO - The Institute of Photonics Sciences, Castelldefels, Barcelona, Spain}

\begin{abstract}
Understanding the fundamental limits of various energy conversion approaches is essential for assessing their efficiency and power output.
In this work, we derive general performance bounds for thermoradiative heat engines that exchange heat radiatively with a cold sink, establishing power-versus-efficiency thermodynamic bounds for several configurations. We find that the performance of these engines is always bounded by that of radiative engines, which harness the thermal radiation emitted by a hot source, making thermoradiative engines inherently less favorable for energy conversion.
By unifying the results of radiative and thermoradiative engines within a common thermodynamic framework, which also encompasses dual-engine configurations that combine both features, this work provides a comprehensive understanding of the thermodynamic limits of radiative energy conversion. Our framework offers general metrics for evaluating specific devices and raises critical questions regarding the relevance of thermoradiative cells for energy production.
\end{abstract}

\maketitle

\section{Introduction}

%\textbf{General background: Heat engines}
\noindent The efficiency of a heat engine operating between a hot source at temperature $T_H$ and a cold sink at temperature $T_C$ is bounded by the Carnot limit $\eta_C = 1 - T_C/T_H$. 
Unfortunately, except for very specific cases~\cite{shiraishi_universal_2016,buddhiraju_thermodynamic_2018,park_nonreciprocal_2022}, operating close to Carnot efficiency leads to vanishing power output. Conversely, operating at maximum power output leads to significantly degraded efficiencies~\cite{curzon_efficiency_1975,van_den_broeck_thermodynamic_2005}, imposing a trade-off between both metrics.
While universal trade-off relations between power and efficiency have been identified~\cite{shiraishi_universal_2016,pietzonka_universal_2018}, they do not provide specific answers for given classes of heat engines.
Due to the increasing importance of developing energy converters across diverse environments, determining the performance bounds of different technologies is essential for evaluating their potential.

%\textbf{General background information: Radiative and thermoradiative heat engines}
Radiative heat engines are non-linear engines that exchange heat in the form of thermal radiation. Conventional radiative heat engines~\cite{giteau_thermodynamic_2023} convert radiation from a hot emitter (\textit{e.g.}, the Sun) while thermoradiative heat engines~\cite{pusch_fundamental_2019} generate power by radiating heat away towards a cold sink (\textit{e.g.}, deep space). Dual systems that combine both types of engines can also be considered~\cite{li_thermodynamic_2020,park_nonreciprocal_2022,legendre_operating_2025}. 

%\textbf{Specific background: Thermoradiative cells - power}
A photovoltaic cell is the canonical example of a conventional radiative heat engine. Its thermoradiative counterpart is known as a \emph{thermoradiative cell}, aiming at extracting energy from the heat radiated by a hot object towards a colder environment. 
Thermoradiative cells have been mainly envisioned in the context of nighttime power generation, whereby an object can be cooled below ambient temperature by radiating towards the cold outer space at night, enabling nighttime photovoltaic power generation~\cite{byrnes_harvesting_2014,deppe_nighttime_2020}. As with solar cells, nighttime thermoradiative cells take advantage of a thermal bath as a ``free" heat source. For solar cells, the heat source is the Sun itself, whereas in the case of nighttime thermoradiative cells, it is the Earth's ambient temperature. 

For both solar cells and nighttime thermoradiative cells, performance can be characterized by a single metric: electrical power generated per unit area. 
However, in applications where energy must be expended to establish and maintain the temperature difference between the thermal reservoirs --- such as thermal energy storage~\cite{datas_latent_2022} or waste heat recovery~\cite{chen_review_2024} --- not only \emph{power output} but also \emph{efficiency} must be taken into account.
In such cases, the trade-off between the two figures of merit is a key consideration in the design and operation of radiative heat engines. 

%\textbf{State of the art - Experiments}
The first experimental demonstration of power generation using a thermoradiative cell was achieved in 2016~\cite{santhanam_thermal--electrical_2016}, followed in 2019 by a demonstration using outer space as the cold sink~\cite{ono_experimental_2019}. More recent experiments have demonstrated significantly larger power densities~\cite{nielsen_thermoradiative_2022} and considered rectenna-based devices~\cite{belkadi_demonstration_2023}. However, their performance to date has remained very limited, especially when compared to their thermophotovoltaic counterparts~\cite{lapotin_thermophotovoltaic_2022,roy-layinde_high-efficiency_2024,giteau_thermodynamic_2024}.
Indeed, all thermoradiative experiments required lock-in detection to generate a measurable signal.

%\textbf{State of the art - Theory}
The performance limits of single- and multi-junction thermoradiative devices have been highlighted in Refs.~\cite{strandberg_theoretical_2015,pusch_fundamental_2019,bohm_fundamental_2025}, both in terms of power and efficiency.
At the same time, the thermodynamic power output limits associated with heat engines radiating heat towards deep space have been addressed extensively in Refs.~\cite{buddhiraju_thermodynamic_2018,li_thermodynamic_2020}.
However, universal power-versus-efficiency trade-off relations for thermoradiative energy conversion, as we previously reported for conventional radiative heat engines~\cite{giteau_thermodynamic_2023}, are currently unknown. 
As a result, it remains an open question whether radiative, thermoradiative, or dual (i.e., a combination of both) approaches offer universally superior performance bounds.

%\textbf{This work}
In this work, we derive the thermodynamic performance limits of thermoradiative heat engines, expressed as a trade-off between power and efficiency.
We consider the nonreciprocal and reciprocal bounds, as well as the endoreversible model, and derive simple analytical expressions when available.
We show that the performance of thermoradiative heat engines is fundamentally bounded by that of conventional radiative heat engines. 
This is especially evident for large temperature differences between the hot and cold baths, while the limits approach each other when the temperatures of the hot and cold baths are similar. Finally, we extend our results to dual engines to provide a complete performance assessment of radiative heat engines.

\section{Framework}

\begin{figure}
\includegraphics[width=.5\textwidth]{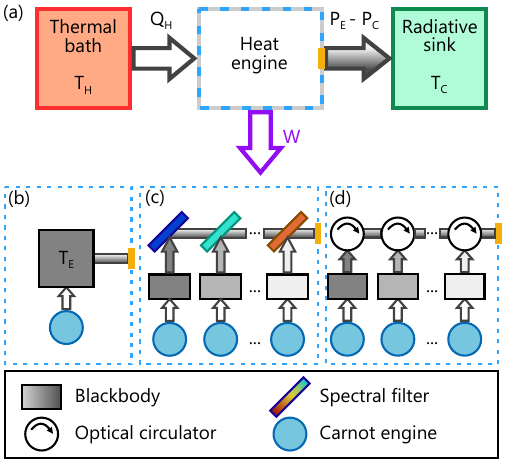}%
\caption{\label{fig:schematic} (a) General representation of a thermoradiative heat engine. (b-d) Different heat engine models considered. (b) Blackbody coupled to a Carnot engine. (c) Infinite nonreciprocal engine for isentropic conversion. (d) Infinite reciprocal engine.
}
\end{figure}

\noindent We consider a heat source at temperature $T_H$ and a heat sink at temperature $T_C$ exchanging power through a heat engine in steady-state. 
The heat source and the converter are in thermal contact, while the converter and the sink exchange heat in the form of thermal radiation.
We call $P_E$ and $P_C$ the power densities emitted by the engine and the sink, respectively. 
The heat engine receives a heat flux $Q_H$ from the source and generates an output power density $W$. The system is schematically represented in Fig.~\ref{fig:schematic}(a). 
In the following, we do not consider near-field effects. We also assume that all photons emitted by the heat engine reach the sink, and vice versa, such that we only need to consider exchanged power \emph{densities} (\textit{i.e.}, power per unit area).

To compare with radiative engines where the engine is in thermal contact with the cold bath~\cite{giteau_thermodynamic_2023}, we define a first figure of merit, $\rho$, as the output power density normalized to the power density emitted by a blackbody emitter at the source temperature $T_H$:

\begin{equation}\label{eq:rho}
    \rho = \frac{W}{\sigma T_H^4},
\end{equation}

\noindent where $\sigma$ is the Stefan-Boltzmann constant. Meanwhile, the second figure of merit is the thermodynamic efficiency $\eta$, for which the power density is normalized to the net heat flux received from the source:
\begin{equation}\label{eq:eta}
    \eta = \frac{W}{Q_H}.
\end{equation}

We note that a thermoradiative engine can be viewed as a radiative engine for which the source and sink temperatures have been exchanged ($T_H<T_C$). Therefore, the work produced by a thermoradiative engine can be calculated from the formulas for conventional radiative energy conversion~\cite{giteau_thermodynamic_2023} by swapping $T_C$ and $T_H$. Nonetheless, the definitions for $\rho$ and $\eta$ should then be modified to reflect the reversed direction of heat flow.

\section{Analytical results}

%\subsection*{Endoreversible model}

\noindent We start by describing a simple endoreversible engine, which consists of a blackbody at temperature $T_E$ coupled to a Carnot engine (Fig.~\ref{fig:schematic}(b)). 
The normalized power output and efficiency, respectively, take the expressions~\cite{de_vos_endoreversible_1993,green_third_2003}:

\begin{equation}\label{eq:rhoBB}
   \rho_E = \left[\frac{T_E^4-T_C^4}{T_H^4}\right]\left[\frac{T_H}{T_E} - 1\right]
\end{equation}

\begin{equation}\label{eq:etaBB}
    \eta_E = 1-\frac{T_E}{T_H}.
\end{equation}

By varying the converter temperature $T_E$ from $T_C$ to $T_H$, Eqs.~\ref{eq:rhoBB}-\ref{eq:etaBB} give the \emph{endoreversible model} for thermoradiative heat engines. In the limit $T_E \to T_H$, the efficiency approaches the Carnot limit, yielding zero power output. The maximum power $\bar{\rho}_E$ is obtained for an engine temperature $T_E$, which satisfies $4 T_E^5-3 T_C T_E^4 - T_C T_H^4 = 0$. 
Interestingly, this is the same condition as for radiative endoreversible engines~\cite{giteau_thermodynamic_2023}.

Endoreversible thermoradiative engines have a strict power bound: $\displaystyle \rho_E \leq 3^3/4^4 \approx 0.11$, approached for $T_C/T_H \to 0$.
When considering radiation from the Earth ($T_H$ = 300 K) into outer space ($T_C$ = 3 K), we find 
$\displaystyle W_B \approx 48.4 \ \mathrm{W.m^{-2}}$, in accordance with Ref.~\cite{buddhiraju_thermodynamic_2018}. A closed-form relation between $\rho_E$ and $\eta_E$ can be derived, which is independent of $T_E$ and depends only on the temperature ratio $T_C/T_H$:

\begin{equation}\label{eq:closedFormBB}
    \rho_E = \frac{\eta_E}{1-\eta_E}\left[\left( 1-\eta_E \right)^4-\left(\frac{T_C}{T_H} \right)^4\right].
\end{equation}

%\subsection*{Transision paragraph with reciprocal bound}

This endoreversible model is not an upper bound for thermoradiative energy conversion. The \emph{reciprocal bound} for thermoradiative heat engines, \textit{i.e.}, the upper bound assuming time-reversal symmetry, is obtained in the limit of an infinite number of endoreversible sub-engines, each converting an infinitesimal part of the incident radiation (Fig.~\ref{fig:schematic}(c)). However, the difference between these two bounds is marginal, so the simple endoreversible model is an appropriate approximation of the reciprocal bound, which is detailed in the Supplemental Material.
The \emph{nonreciprocal bound} for thermoradiative heat engines, which is the absolute limit for thermoradiative energy conversion, is presented below.

%\subsection*{Nonreciprocal bound}

We consider a thermoradiative heat engine performing \emph{isentropic} energy conversion, similar to the approach considered by Landsberg for solar energy conversion~\cite{landsberg_thermodynamic_1980,green_third_2003}. We consider that the heat engine is at temperature $T_E$, emitting a power density $P_E = \sigma T_E^4$. To approach an isentropic conversion process, the converter must combine an infinite number of endoreversible sub-engines coupled via nonreciprocal optical elements, using either optical circulators~\cite{ries_complete_1983} or non-reciprocal emitters that break Kirchhoff's law of thermal radiation~\cite{shayegan_direct_2023}.
In that configuration, each sub-engine receives radiation from the colder sub-engines (to the right) and emits towards the hotter sub-engines (to the left)~\cite{ries_complete_1983,buddhiraju_thermodynamic_2018,park_reaching_2022} (see Fig.~\ref{fig:schematic}(d)). The heat and power flux can be calculated from power and entropy conservation:

\begin{align}
    Q_H &= \frac{4}{3} \sigma T_H\left( T_E^3-T_C^3\right) \\
    W     &= \frac{4}{3} \sigma T_H\left( T_E^3-T_C^3\right) - \sigma \left( T_E^4 -T_C^4 \right),
\end{align}

\noindent leading to the following figures of merit:

\begin{equation}\label{eq:rhoN}
    \rho_N = \frac{4}{3} \left( \frac{T_C}{T_H} \right)^3 \left[\left(\frac{T_E}{T_C}\right)^3 -1 \right] - \left( \frac{T_C}{T_H} \right)^4 \left[\left(\frac{T_E}{T_C}\right)^4 -1 \right]
\end{equation}

\begin{equation}\label{eq:etaN}
    \eta_N = 1 - \frac{3}{4}\frac{T_C}{T_H}\frac{\left(\frac{T_E}{T_C}\right)^4-1}{\left(\frac{T_E}{T_C}\right)^3-1}.
\end{equation}

By adjusting the engine temperature $T_E>T_C$ such that $\eta_N$ is varied from 0 to $\eta_C$, the power-efficiency relation can be determined from Eqs.~\ref{eq:rhoN}-\ref{eq:etaN}, leading to the \emph{nonreciprocal bound} for thermoradiative heat engines.
In the Supplementary Material, we present a closed-form expression (independent of $T_E$) relating $\eta_N$ and $\rho_N$ and
justify why the relationship is only a function of the temperature ratio $T_C/T_H$.

For $T_E \to T_C$, the efficiency tends to the Carnot limit $\eta_C$ as the power density vanishes. By contrast, for $T_E = T_H$, the power output is maximized, leading to:

\begin{align}\label{eq:PL}
    \bar{\rho}_N &= \frac{1}{3} - \left(\frac{T_C}{T_H}\right)^3 \left[ \frac{4}{3}  - \frac{T_C}{T_H}  \right] \\
    \bar{\eta}_N &= \frac{1}{4} - \frac{3}{4} \frac{\left(\frac{T_C}{T_H}\right)^3}{1+\frac{T_C}{T_H} + \left(\frac{T_C}{T_H}\right)^2}.
\end{align}

Introducing nonreciprocity, the power that can be generated through radiation from the Earth (300 K) into the cold outer space (3 K) is $W_N \approx 153 \ \mathrm{W.m^{-2}}$, as calculated by Buddhiraju \textit{et al.}~\cite{buddhiraju_thermodynamic_2018}. While operating at $T_E>T_H$ may lead to positive power output, it is always sub-optimal as both efficiency and power density are lower than for $T_E = T_H$. Interestingly, from Eq. \ref{eq:PL}, there is an absolute limit $\bar{\rho}_N \leq 1/3$ on the normalized power output, corresponding to an efficiency at maximum power $\bar{\eta}_N \leq 1/4$. The equality condition is approached as $T_C/T_H \to 0$. This power limit can be understood as the free energy or exergy of radiation into space (related to the $4/3$ factor in the entropy of radiation). It was first identified by Würfel, although he initially presented it as a paradox violating the second law of thermodynamics~\cite{wurfel_chemical_1982}.

%\subsection*{Linear approximations}
Linear approximations of the endoreversible model and nonreciprocal bound close to the Carnot limit ($T_E \to T_C$), detailed in Eqs.~9-12 of the Supplemental Material, lead to:

\begin{align}
    \rho_E &\approx 4 \left(\frac{T_C}{T_H}\right)^2 \eta_C (\eta_C-\eta_E) \label{eq: rhoB_approx} \\
    \rho_N &\approx 8 \left(\frac{T_C}{T_H}\right)^2 \eta_C (\eta_C-\eta_N).\label{eq: rhoN_approx}
\end{align}

Furthermore, as $T_C/T_H \to 1$, a second-order expansion in $\eta$ leads to:
\begin{align}
    \rho_E &\approx 4 \eta_E (\eta_C-\eta_E) \label{eq:rhoB_parabolic} \\
    \rho_N &\approx 8 \eta_N (\eta_C-\eta_N). \label{eq:rhoN_parabolic}
\end{align}

These expressions are identical to those found for conventional radiative heat engines~\cite{giteau_thermodynamic_2023}. 
This generalizes the previous observation (both theoretical~\cite{strandberg_theoretical_2015} and experimental~\cite{nielsen_thermoradiative_2022}) that radiative and thermoradiative cells tend to offer similar performance for small temperature differences.

%\section*{Illustrations}
\begin{figure}
\includegraphics[width=.5\textwidth]{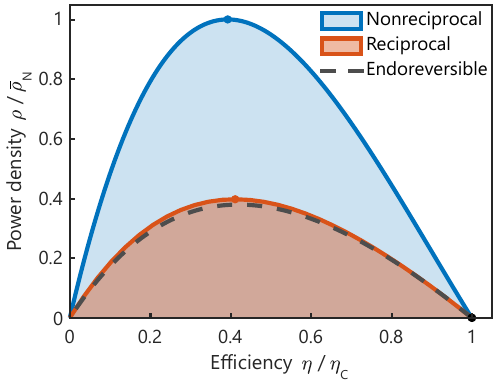}%
\caption{\label{fig:fig2} Thermodynamic performance limits of thermoradiative energy conversion: maximum power output $\rho$ (normalized to the nonreciprocal power limit $\bar{\rho}_N$) as a function of the efficiency $\eta$ (normalized to the Carnot efficiency $\eta_C$) for the nonreciprocal bound (blue area), reciprocal bound (red area) and endoreversible model (dashed black line).
The heat source temperature is $T_H = 600 \ \mathrm{K}$ and the sink temperature is $T_C = 300 \ \mathrm{K}$.}
\end{figure}

\section{Power-efficiency bounds}

%\subsection*{Fig 2: comparing the bounds}

\noindent We present in Fig.~\ref{fig:fig2} the power-efficiency relation for the nonreciprocal (Eqs.~\ref{eq:rhoN}-\ref{eq:etaN}) and reciprocal (see Supplemental Material) bounds as well as the endoreversible model (Eqs.~\ref{eq:rhoBB}-\ref{eq:etaBB}) for a temperature ratio $T_C/T_H = 0.5$.
The reciprocal limit and endoreversible model are similar, though we note that they are not as close as in the case of conventional radiative heat engines~\cite{giteau_thermodynamic_2023}. At the same time, a significant performance improvement can be achieved by breaking reciprocity. Indeed, for any given efficiency $\eta$ and temperature ratio $T_C/T_H$, the power output limit of nonreciprocal systems is always at least twice as large as for reciprocal systems, as justified in the Supplemental Material.

%\subsection*{Fig 3: bounds for different temperatures}

\begin{figure}
\includegraphics[width=.5\textwidth]{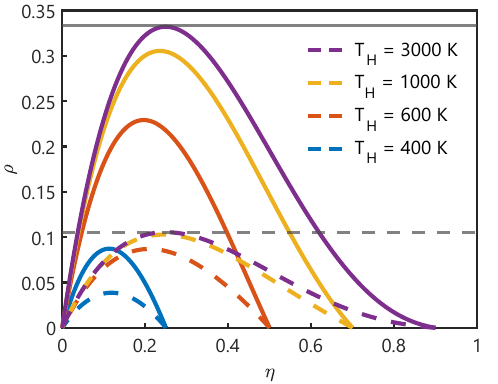}%
\caption{\label{fig:fig3} Maximum power $\rho$ as a function of the efficiency $\eta$ for different temperature ratios for the nonreciprocal bound (solid lines) and the endoreversible model (dashed lines).
The absolute maximum achievable output power for each paradigm is indicated by the gray lines.}
\end{figure}

Next, we show in Fig.~\ref{fig:fig3} the power-efficiency relation derived for different temperature ratios, focusing on the endoreversible and nonreciprocal cases. As $T_H \gg T_C$, the performance progressively approaches the maximum achievable power (discussed earlier), shown in dashed and solid gray lines, respectively. We also observe that $\partial \rho / \partial \eta \to 0 $ close to the Carnot limit, a feature we can infer from Eqs.~\ref{eq: rhoB_approx}-\ref{eq: rhoN_approx}.

\section{Comparison with radiative and dual heat engines}

%\subsection*{Introducing dual engines}

\begin{figure*}
\includegraphics[width=\textwidth]{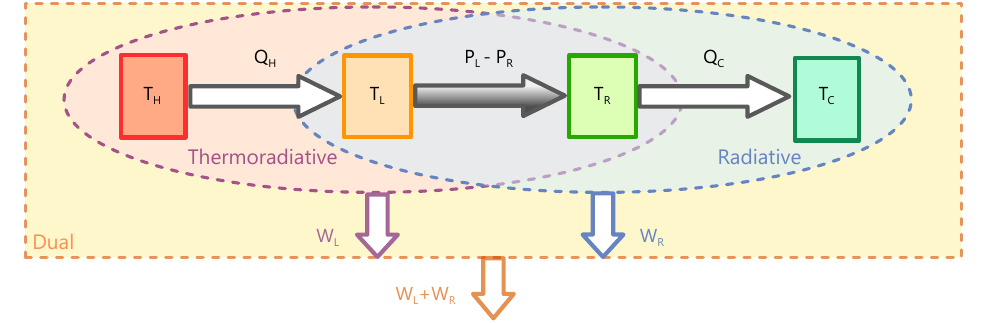}%
\caption{\label{fig:fig4}
Illustration of a dual radiative engine and its decomposition into radiative and thermoradiative sub-parts.
}
\end{figure*}

\noindent So far, we presented the performance of thermoradiative heat engines, where the engine is in contact with the hot source. In our previous work in Ref.~\cite{giteau_thermodynamic_2023}, we considered the performance limits of radiative heat engines where the engine is in contact with the cold sink. To complete this framework, here, we also consider the possibility of implementing dual radiative heat engines that generate power on both the hot and cold side, as previously considered in Refs.~\cite{li_thermodynamic_2020,park_nonreciprocal_2022,legendre_operating_2025}. With respect to Fig.~\ref{fig:fig4}, we consider a dual engine with a left engine at temperature $T_L$ (hot side) generating work $W_L$, and a right engine at temperature $T_R$ (cold side) generating work $W_R$. We impose the inequalities $T_H \geq T_L \geq T_R \geq T_C$ such that $W_L$ and $W_R$ are both positive.
The total power output is $W = W_L + W_R$. With these notations, the definitions of normalized power and efficiency of Eqs.~\ref{eq:rho}-\ref{eq:eta} remain identical for dual engines. In analogy with the previous sections, we consider two cases: the endoreversible model, which considers blackbodies coupled to Carnot engines, and the nonreciprocal bound, assuming no entropy generation on either side. The formulas are obtained by combining the results of thermoradiative engines derived in this paper with those of conventional radiative heat engines from Ref.~\cite{giteau_thermodynamic_2023}. 

%\subsection*{Endoreversible model}

For the endoreversible case, the power and efficiency of dual engines take the expressions:

\begin{align} 
    \rho_E^d &= \left(\frac{T_R}{T_H} \right)^3 \left[ \left(\frac{T_L}{T_R} \right)^4 -1 \right] \left[ \frac{T_R}{T_L} - \frac{T_C}{T_H} \right] \label{eq:rhoDE}\\
    \eta_E^d &= 1-\frac{T_L}{T_R} \frac{T_C}{T_H}. \label{eq:etaDE}
\end{align}

We verify $\eta_E^d \leq \eta_C$ from the constraint $T_L \geq T_R$. For a given efficiency, from Eq.~\ref{eq:etaDE}, the ratio $T_L/T_R$ is fixed, hence the power output is an increasing function of $T_R$ (see Eq.~\ref{eq:rhoDE}). Therefore, the optimal configuration leads to:

\begin{align}
    T_L &= T_H \\
    \eta_E^d &= 1-\frac{T_C}{T_R}. 
\end{align}

These expressions coincide with the endoreversible model for conventional radiative engines ($W_L=0$). Thereby, adding a thermoradiative engine on the hot side yields no performance improvement in the endoreversible case. 
We argue that this conclusion extends to the reciprocal bound, for the following reasons. First, the argument derived for a single engine can be made on a per-photon-energy basis. Second, the results obtained for the endoreversible model are almost identical to those of reciprocal dual engines presented by Park et al. in Ref.~\cite{park_nonreciprocal_2022}, Fig.~3(d), when considering the same temperature conditions. 

As a result, \emph{the performance of any reciprocal system is practically bounded by the endoreversible model for conventional radiative engines.}
This implies that employing thermoradiative cells in place of --- or even in addition to --- thermophotovoltaic cells cannot be justified from a performance standpoint. Rather, efforts are better directed towards approaching the endoreversible limit through a combination of multi-junction thermophotovoltaic cells~\cite{lapotin_thermophotovoltaic_2022} and minimized sub-bandgap heat transfer~\cite{roy-layinde_high-efficiency_2024}.

We note that the performance could be substantially enhanced by allowing $W_L < 0$. This regime corresponds to thermophotonics, where the hot-side engine operates as a light-emitting diode rather than a thermoradiative cell~\cite{legendre_operating_2025}. However, this configuration remains experimentally challenging due to the exceptionally high radiative efficiency required of the LED, and no conclusive demonstration has been reported to date.

%\subsection*{Nonreciprocal bound}

Next, we consider the case of nonreciprocal dual engines, where the performance bounds are:

\begin{align}
    \rho_N^d &= \frac{4}{3} \left[1-\frac{T_C}{T_H} \right] \left[\left(\frac{T_L}{T_H}\right)^3 - \left(\frac{T_R}{T_H}\right)^3 \right] \label{eq:rhoND} \\
    \eta_N^d &= \eta_C \label{eq:etaND}.
\end{align}

This nonreciprocal dual case was previously investigated in Refs.~\cite{buddhiraju_thermodynamic_2018,park_nonreciprocal_2022}, demonstrating the peculiar fact that a nonreciprocal dual engine can produce finite power at Carnot efficiency. In this idealized case, there is therefore no trade-off between power and efficiency. This was shown to be compatible with thermodynamic principles, although it implies diverging fluctuations~\cite{pietzonka_universal_2018}. Assuming both engines generate positive work, maximizing power yields $T_L = T_H$ and $T_R = T_C$, leading to:

\begin{equation}
    \rho_{N,max}^d = \frac{4}{3} \left[1-\frac{T_C}{T_H} \right] \left[1-\left(\frac{T_C}{T_H}\right)^3 \right] \label{eq:rhomaxND},
\end{equation}

\noindent which is the sum of the maximum power generated by nonreciprocal radiative and thermoradiative engines. Since a dual nonreciprocal heat engine simultaneously harvests radiation from both the hot and cold reservoirs, it can extract more available work than either engine alone. Notably, the maximum normalized power output $\rho_{N,max}^d$ can exceed unity, approaching 4/3 for $T_C/T_H \to 0$, in accordance with the exergy of thermal radiation.

%\subsection*{Illustration}

\begin{figure*}
\includegraphics[width=\textwidth]{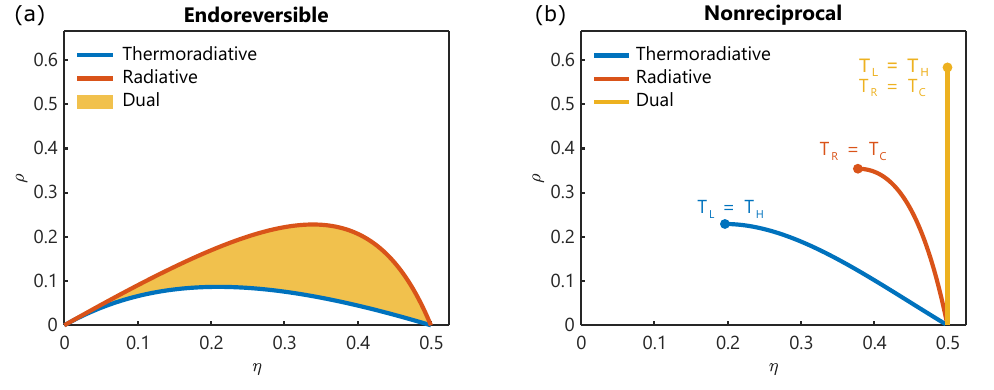}%
\caption{\label{fig:fig5}
Power versus efficiency characteristics of (a) endoreversible and (b) nonreciprocal heat engines when considering engines exclusively on the hot side (thermoradiative) in blue, on the cold side (radiative) in red, or on both sides (dual) in yellow, for a temperature ratio $T_C/T_H = 0.5$.}
\end{figure*}

All aforementioned results are summarized in Table~\ref{tab:formulas}, and the performance is illustrated in Fig.~\ref{fig:fig5} for the endoreversible (panel a) and nonreciprocal (panel b) cases, for $T_C/T_H = 0.5$. In the Supplemental Material, we also consider the cases $T_C/T_H = 0.1$ and $T_C/T_H = 0.9$. Since in the case of endoreversible dual engines $T_L$ and $T_R$ can be varied independently, the dual results in panel (a) cover an area instead of being a curve.

As stated above, in the endoreversible model, there is no benefit to operating a dual engine compared to a purely radiative one. Accordingly, assuming reciprocal heat transfer, implementing thermoradiative cells provides no advantage. By contrast, as shown in panel (b) and inferred from Eqs.~\ref{eq:etaND}-\ref{eq:rhomaxND}, nonreciprocal radiative engines benefit strongly from dual engine operation, enabling finite power generation even at the Carnot efficiency limit.

Overall, the most crucial observation is that \emph{for any given temperature ratio and efficiency, conventional radiative heat engines consistently yield higher power output than thermoradiative ones.}
The difference is particularly striking for large temperature differences, whereas for $T_C/T_H \to 1$, heat transfer becomes linear with temperature such that both models converge towards the same parabolic behavior (see Eqs.~\ref{eq:rhoB_parabolic}-\ref{eq:rhoN_parabolic}). 

\begin{table*}
    \centering
    \begin{tabular}{|c|c|c|}
    \hline
         & Endoreversible & Nonreciprocal\\
         \hline
       Radiative & \begin{tabular}{c}
            $\rho_E = \left[1-\left(\frac{T_E}{T_H}\right)^4\right]\left[1-\frac{T_C}{T_E}\right] $  \\
            \vspace{2mm}
            $\eta_E= 1-\frac{T_C}{T_E} $
       \end{tabular}

        & \begin{tabular}{c}
            $\rho_N = 1-\frac{4}{3}\frac{T_C}{T_H}-\left(\frac{T_E}{T_H}\right)^4\left[1-\frac{4}{3}\frac{T_C}{T_E}\right]$  \\
            \vspace{2mm}
            $\eta_N = \frac{\rho_N}{1 - \left(\frac{T_E}{T_H} \right)^4}$
       \end{tabular}\\
\hline
       Thermoradiative & \begin{tabular}{c} $\rho_E = \left[\frac{T_E^4-T_C^4}{T_H^4}\right]\left[\frac{T_H}{T_E} - 1\right] $ 
       \vspace{2mm}\\
       $ \eta_E = 1-\frac{T_E}{T_H}$ 
       \end{tabular}
       
       & \begin{tabular}{c} $\rho_N = \frac{4}{3} \left( \frac{T_C}{T_H} \right)^3 \left[\left(\frac{T_E}{T_C}\right)^3 -1 \right] - \left( \frac{T_C}{T_H} \right)^4 \left[\left(\frac{T_E}{T_C}\right)^4 -1 \right] $
       \vspace{2mm}\\
       $ \eta_N =1 -\frac{3}{4}\frac{T_C}{T_H}\frac{\left(\frac{T_E}{T_C}\right)^4-1}{\left(\frac{T_E}{T_C}\right)^3-1}.$
       \end{tabular}\\
       \hline
       Dual & \begin{tabular}{c}
          $\rho_E = \left(\frac{T_R}{T_H} \right)^3 \left[ \left(\frac{T_L}{T_R} \right)^4 -1 \right] \left[ \frac{T_R}{T_L} - \frac{T_C}{T_H} \right] $
                 \vspace{2mm}\\
             $ \eta_E = 1-\frac{T_L}{T_R} \frac{T_C}{T_H} $
       \end{tabular} & \begin{tabular}{c}$\rho_N = \frac{4}{3} \left[1-\frac{T_C}{T_H} \right] \left[\left(\frac{T_L}{T_H}\right)^3 - \left(\frac{T_R}{T_H}\right)^3 \right]$
       \vspace{2mm}\\
       $\eta_N= \eta_C$  \end{tabular}\\
       \hline
    \end{tabular}
    \caption{Summary of the power-efficiency relations for different kinds of radiative engines, considering the endoreversible model and the nonreciprocal limit.}
    \label{tab:formulas}
\end{table*}

\section{Conclusion}

In this work, we derived performance bounds for thermoradiative energy conversion, expressed in terms of the maximum power density achievable at a given efficiency. We demonstrated that, for any efficiency, radiative energy converters consistently generate more power when the engine operates on the cold side.
In practical terms, this implies that \emph{optimal thermophotovoltaic devices will always outperform optimal thermoradiative devices}. This conclusion holds even before considering that thermoradiative cells operate on the \emph{hot} side, where elevated temperatures exacerbate non-radiative recombination, thereby further limiting efficiency.

For small temperature differences between hot and cold reservoirs, radiative and thermoradiative energy conversion exhibit similar performance limits. However, in this regime, thermoelectric devices are typically more suitable than radiative heat engines~\cite{tedah_thermoelectrics_2019}. As the temperature difference increases, thermoradiative performance saturates and becomes dominated by that of radiative engines. Introducing a combination of engines on the hot and cold side offers potential gains, but only when nonreciprocal elements are introduced. In that regard, promising results have been obtained by breaking Kirchhoff's law of thermal radiation~\cite{shayegan_direct_2023}. Still, this remains to be translated to developments on nonreciprocal energy converters, which require extremely efficient implementations of nonreciprocity for specific spectral and angular ranges.
As a result, \emph{all practical radiative engines can be considered bounded by the radiative endoreversible model}, which therefore serves as a critical reference point for evaluating device performance~\cite{giteau_thermodynamic_2024}. Besides nonreciprocity, the only ways to surpass this limit involve super-Planckian radiative exchange, for example, using thermophotonics~\cite{harder_thermophotonics_2003}, near-field effects~\cite{laroche_near-field_2006,pascale_perspective_2023}, or a combination of both~\cite{legendre_gaas-based_2022}.

\section*{Acknowledgments}
G.T.P. acknowledges the support of the Spanish MICINN (PID2021-125441OA-I00, PID2020-112625GB-I00, and CEX2019-000910-S), the European Union (fellowship LCF/BQ/PI21/11830019 under the Marie Skłodowska-Curie Grant Agreement No. 847648), Generalitat de Catalunya (2021 SGR 01443), Fundació Cellex, and Fundació Mir-Puig. M.G. acknowledges financial support from the Severo Ochoa Excellence Fellowship. M.F.P. acknowledges support from the Optica Foundation 20th Anniversary Challenge Award. M.F.P. and G.T.P. received the support of fellowships from “la Caixa” Foundation (ID 100010434). The fellowship codes are LCF/BQ/PI23/11970026 and LCF/BQ/PI21/11830019.

\section*{References}

\bibliography{biblio.bib}

\end{document}

\title{Fundamental limitations of thermoradiative energy conversion \\ Supplemental Material}

\author{Maxime Giteau}
 \affiliation{CNRS, Laboratoire PROcédés, Matériaux et Energie Solaire (PROMES), UPR 8521, Odeillo, France}
 \affiliation{ICFO - The Institute of Photonics Sciences, Castelldefels, Barcelona, Spain}
\author{Michela F. Picardi}%
\author{Georgia T. Papadakis}%
\affiliation{ICFO - The Institute of Photonics Sciences, Castelldefels, Barcelona, Spain}

\maketitle

\section*{Reciprocal bound}

Here, we derive the performance bounds of reciprocal thermoradiative heat engines, following the steps taken in Ref.~\cite{giteau_thermodynamic_2023} for the conventional radiative case.
We assume the converter consists in an infinite set of blackbodies at temperature $T_E(E)$, each converting the radiation from a spectral element $[E,E+dE]$ through a Carnot engine (Fig.~1(c) in the manuscript). We define the photon occupation numbers 

\begin{align}
    n_E (E) &= \frac{1}{\exp[E / k T_E(E)]-1} \\
    n_C (E) &= \frac{1}{\exp[E / k T_C]-1},
\end{align}
\noindent as well as $\displaystyle \Delta n (E) = n_E(E) - n_C(E)$.
The power generated by each endoreversible engine is:

\begin{equation}\label{eq:dPMC}
    dW = \frac{2 \pi}{h^3 c^2} E^3  \Delta n (E) \left(\frac{T_H}{T_E(E)} - 1 \right) dE.
\end{equation}

By summing the power generated by all engines, i.e., by integrating $dW$, we obtain

\begin{equation}\label{eq:rhoMC}
    \rho_{R} = \frac{15}{\pi^4 k^4 T_H^4} \int_0^\infty E^3 \Delta n(E) \left[ \frac{T_H}{T_E(E)} -1 \right] dE
\end{equation}

\begin{equation}\label{eq:etaMC}
    \eta_{R} = 1 - \frac{ \int_0^\infty E^3 \Delta n (E) dE }
    { \int_0^\infty E^3 \frac{T_H}{T_E(E)} \Delta n (E) dE}.
\end{equation}

For a given $\eta_R$, $T_E$ functions must satisfy, according to Eq.~\ref{eq:etaMC}: 

\begin{equation}
    \int_0^\infty E^3 \left[\frac{1}{T_E(E)} - \frac{1}{T_F} \right] \Delta n (E) dE = 0,
\end{equation}

\noindent where $\displaystyle T_{F} = T_H (1-\eta_R)$, verifying $\displaystyle T_C \leq T_F \leq T_H$.
A trivial solution is for all blackbodies to be at the same temperature $T_E = T_F$, leading to the endoreversible thermoradiative model presented in the main text.
However, this is not the upper bound for reciprocal radiative energy conversion, as the temperature profile can be refined to improve performance.

We want to determine the temperature profile $T_E(E)$ that maximizes the power output for any given efficiency.
First, we calculate the absolute maximum power for reciprocal radiative energy converters by finding the temperature $T_E$ that maximizes $dW$ (Eq.~\ref{eq:dPMC}) for each photon energy.
For radiation from the Earth into outer space, we recover $W_R \approx 55.0 \ \mathrm{W.m^{-2}}$, in accordance with ref.~\cite{buddhiraju_thermodynamic_2018}.
More generally and contrary to Ref.~\cite{giteau_thermodynamic_2023}, we find from Eqs.~\ref{eq:rhoMC}-\ref{eq:etaMC} that, for a given efficiency $\eta_R$, maximizing the power density is equivalent to \emph{maximizing} the emitted power density 

\begin{equation}
    P_E \propto \int_0^\infty E^3 n_E(E) dE.
\end{equation}

\begin{figure} [h]
\includegraphics[width=\columnwidth]{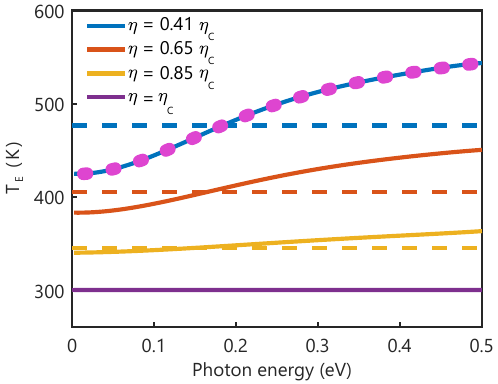}%
\caption{\label{fig:figS1} Engine temperature $T_E$ as a function of the photon energy for the reciprocal bound (solid lines) compared to the fixed temperature of the endoreversible model (dashed lines) for different efficiency points. The blue line corresponds to the temperature profile for the maximum power point $\bar{\rho}_R$. The source temperature is $T_H= 600 \ \mathrm{K}$ and the sink temperature is $T_C = 300 \mathrm{K}$. The pink dots show the point-by-point optimization that leads to maximum power.}
\end{figure}

To find the optimal solutions, we model $T_E(E)$ as a polynomial and adjust its coefficients to minimize the power output.
We show in Fig.~\ref{fig:figS1} the temperature profiles $T_E(E)$ that lead to the reciprocal bound of thermoradiative engines for several efficiencies (solid lines), for the conditions shown in Fig.~2 of the article ($T_H = 600 \ \mathrm{K}$ and $T_C = 300 \ \mathrm{K}$).
These temperature profiles are obtained by fitting the coefficients of a 7th order polynomial to maximize the power for a given efficiency.
They are compared to the blackbody temperatures ($T_E = T_F$) for the endoreversible model (dashed lines).
The maximum power obtained by polynomial fitting (blue line) perfectly matches the energy-by-energy maximization of the output power (pink dots), validating the optimization approach.

\section*{Closed-form expression for the nonreciprocal bound}

We define $x=T_C/T_H$ and $y = T_E/T_C$.
The efficiency relation for nonreciprocal thermoradiative heat engines (Eq.~9 in the main text) can be re-expressed as:

\begin{equation}
    \left(1-\eta_N \right) \left(1 + y + y^2 \right) = \frac{3}{4}x\left(1 + y + y^2 + y^3 \right).
\end{equation}

Therefore, for a given $x$, $\eta_N$ and $y$ are related through a 3rd-order polynomial equation. Its discriminant is negative, confirming there is a single real root of the form $y = f(\eta_N,x)$, given at the end of this document. 
We can then inject this root into the power expression (Eq.~8 in the main text) to obtain the closed-form expression:

\begin{equation}
    \rho_N = \frac{4}{3}x^3 \left(y^3-1\right) - x^4 \left(y^4-1\right).
\end{equation}

\section*{Power ratio between nonreciprocal and reciprocal limits}

For $T_E \to T_C$, the efficiency and power of nonreciprocal systems follow, in first-order approximation:

\begin{align}
    \eta_N &\approx \eta_C - \frac{1}{2} \frac{T_C}{T_H} \left(\frac{T_E}{T_C}-1 \right) \\
    \rho_N &\approx 4 \left(\frac{T_C}{T_H}\right)^3 \left(\frac{T_E}{T_C}-1 \right) \left(1-\frac{T_C}{T_H}\right),
\end{align}

\noindent while for the endoreversible model, we have:

\begin{align}
    \eta_{E} &\approx \eta_C - \frac{T_C}{T_H} \left(\frac{T_E}{T_C}-1 \right) \\
    \rho_{E} &\approx 4 \left(\frac{T_C}{T_H}\right)^3 \left(\frac{T_E}{T_C}-1 \right) \left(1-\frac{T_C}{T_H}\right).
\end{align}

\begin{figure}[h]
\includegraphics[width=\columnwidth]{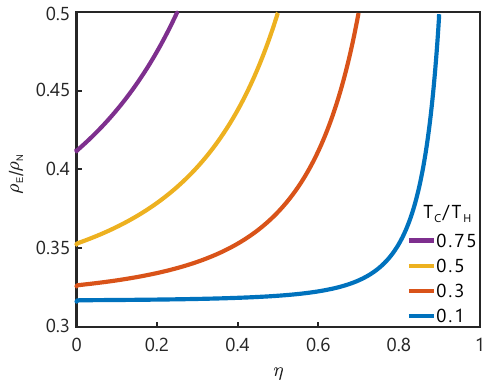}%
\caption{\label{fig:figS2} Ratio between the power outputs for the endoreversible model $\rho_E$ and the nonreciprocal bound $\rho_N$ as a function of the efficiency $\eta$, for different temperature ratios.}
\end{figure}

Therefore, as the converter temperature increases from $T_C$, the power for both the endoreversible model and the nonreciprocal limit increase similarly, while the efficiency drops twice as fast for the endoreversible case.
As a result, the ratio $\rho_E/\rho_N$ should take a value of 0.5 for $\eta \to \eta_C$ and decrease for lower efficiencies, as illustrated for different temperature ratios in Fig.~\ref{fig:figS2}.
We observed numerically that this result also holds when considering the ratio between reciprocal and nonreciprocal limits.

%\section*{Comparison between radiative, thermoradiative, and dual for small and large temperature ratios}

\begin{figure*} [h]
\includegraphics[width=\textwidth]{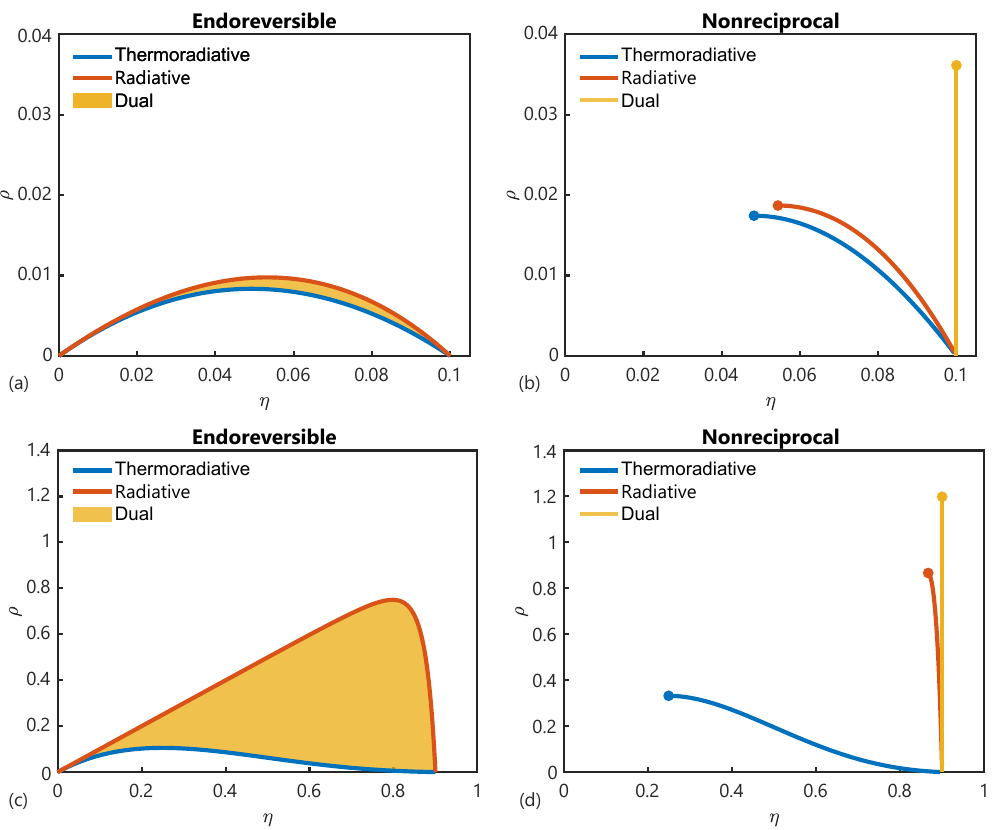}%
\caption{\label{fig:figS3} Power versus efficiency characteristics of (a, c) endoreversible and (b, d) nonreciprocal radiative heat engines when considering engines exclusively on the hot side (thermoradiative) in blue, on the cold side (radiative) in red or on both sides (dual) in yellow. (a-b) are calculated for $T_C/T_H = 0.9$ (similar temperatures), while (c-d) correspond to $T_C/T_H = 0.1$ (extreme temperature difference).}
\end{figure*}

\begin{figure*} [h]
\includegraphics[width=\textwidth]{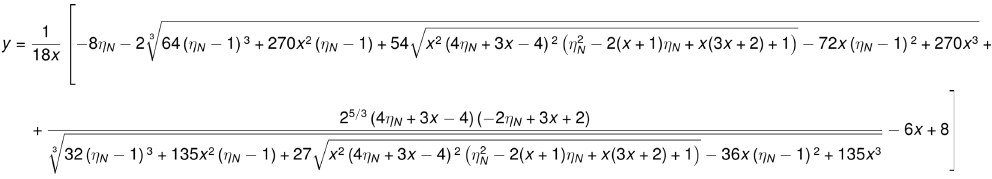}%
\end{figure*}

\section*{References}

\bibliography{biblio.bib}